\def\be{\begin{equation}}
\def\ee{\end{equation}}
\documentstyle[aps,twocolumn,prl]{revtex}

\begin{document}
\draft
\preprint{BIXletter}
\title{ The (In)stability of the Bianchi IX Dynamics:\\
The Geodesic Deviation Equation in the Finsler Spaces}
\author{M. Di Bari$^{1,*}$ and P. Cipriani$^{2,\dagger}$}
\address{$ ^1$Dipartimento di Fisica, Universit\`a di Parma, viale delle Scienze, - 43100 - Parma.\\
$ ^2$Dipartimento di Fisica, Universit\`a ``Tor Vergata", via della Ricerca Scientifica, 1 
- 00133 - Roma.}
\date{June 16, 1998}
\maketitle
\begin{abstract}
We explore the dynamical stability of the minisuperspace Hamiltonian of the Bianchi IX 
cosmological models, giving a gauge-invariant and unapproximated description of the full 
continuous dynamics, achieved through a geometrical description of the equations of 
motion in the framework of the theory of Finsler Spaces.
The numerical integrations of the geodesics and geodesic deviation equations
show clearly the absence of any ``traditional" signature of Chaos, 
while suggesting a chaotic scattering dynamics scenario. 
\end{abstract}
\narrowtext
The characterization of Chaos 
in the framework of General Re\-la\-ti\-vi\-ty is still an open question, receiving
recently an increasing attention \cite{Hobill_book}.
The issue is related to the well known ambiguities hidden in any gauge theory, 
and in particular to 
the freedom in the choice of the coordinate system. A gauge invariant description 
of Chaos must be 
invariant under any such choice, in particular with respect 
to rescaling of the ``time" parameter.
The problem is mainly related to the choice of an invariant {\it internal} time \cite{Prigo} 
which
must be adopted in describing the evolution of a general relativistic dynamical system. 
The applications to the Bianchi IX dynamics of the {\it standard} approaches to Chaos 
borrowed from classical theory of dynamical systems have, in general, revealed their
inadequacy to satisfy this requirement. 
This has been outlined in the past \cite{Berger}, in particular for what concerns 
the dependence of the Lyapunov characteristic numbers on the choice of the time variable.

The main works devoted to the study of Bianchi IX dynamical behaviour are essentially of two
kinds. On one hand the first studies of Belinski et al. \cite{Bel} and the subsequent 
papers \cite{Barrow}, till the last of Cornish and Levin \cite{Cornish} are based on a 
reduction of the continuous flow to discrete maps (i.e. Gauss and Farey maps) 
which mark the more relevant events in the evolution. 
In this case the {\sl time} is a discrete variable which increases once a particular 
event occurs, i.e. when there is a bounce on the potential walls in the minisuperspace 
framework \cite{Misner} or, in the other picture, a crossing to a new Kasner epoch and/or 
era \cite{Bel}. These events provide a natural clock for the evolution of the system.
The discrete approximation of dynamics by maps is usually 
adopted also in classical mechanics \cite{maps}, where the 
definition of time is not questioned and the interval between two successive increments
in the maps' variables, is roughly constant.
This is not the case for Bianchi IX models, as 
the lapse of time depends on the definition of time itself, unless an invariant choice
is adopted.
In \cite{Cornish}, with the aim of giving an {\sl intrinsic} description and relying on 
the observer-independent properties of fractals in phase space, 
the authors argue a chaotic behaviour of the full dynamics by calculating the 
strange repellor's Lyapunov dimension, the topological entropy and the 
multifractal dimensions of the Farey map.\\
However, even if the maps reproduce the main characteristic of the flow in the Kasner regime, 
they constitute only a discrete approximation of the actual field equations
for the Bianchi IX models; so, it is reasonable that some efforts have been devoted 
to analyse the continuos dynamical flow.
Within the framework of a gauge-invariant description of motions and of their possibly 
chaotic properties, the geometrical description of the dynamics offers a natural tool, 
as it gives a metric in terms of which also an invariant time variable is defined.
The first attempt towards this kind of approach dates back to Chitre \cite{Chitre}, 
which reduced the motion to a geodesic flow on a hyperbolic
manifold, in order to advocate the mixing properties of geodesics on a compact Riemannian 
manifold with constant negative curvature \cite{Anosov}.
However, some approximations need to be assumed, which are ultimately equivalent
to those involved for the maps.
In the hope of avoiding any approximation, some authors \cite{Szydlowsky} adopted 
the Jacobi metric which derives from the Maupertuis principle and which has been 
exploited in classical mechanics, in order to understand the sources of Chaos in
dynamical systems\cite{commento1}.
The internal time is given in this case by
\be
ds_J=\sqrt{2}\left[ E-U(x^i) \right]dt
\ee
where $E$ is the total energy and $U(x^i)$ is the potential energy depending 
on the coordinates $\{x^i\}$ alone.
The manifold is again Riemannian but the curvature is not constant 
and negative.
The features of the motion are now ascribed to the average behaviour of 
the Ricci curvature. A clear detection of a predominantly negative Ricci curvature,
which goes to $-\infty$ whenever there is a bounce, 
would indicate the presence of instability,
at least in a local sense rather than as a global property of the motion \cite{Biesiada}.
Neverthless, in a general relativistic context, it is $E=0$, because of the covariant 
character of the theory, so 
$T \equiv - U$ and $ds_J = \sqrt{2} T dt = - \sqrt{2} U  dt$.
Therefore the kinetic term is not positive definite and, even imposing by hand that 
$ds_J=\sqrt{2} \vert T \vert dt = \sqrt{2} \vert U\vert dt$, 
the relation between the time $t$ and the ``Jacobi parameter" $s_J$ becomes 
meaningless when $T = - U = 0$, which leads to singularities in the sectional, 
Ricci and scalar curvatures. So, although a Jacobi metric can be introduced without 
any approximation, in order to analyse the curvature properties, it is necessary {\sl to cut} 
the trajectory when these singularities occur.
For this reason, the Jacobi metric cannot be properly used \cite{Burd}.
As noticed in \cite{Aquila2}, here the zero kinetic energy surface no longer constitutes 
the external boundary of the region allowed to motion in configuration space, 
so it can be crossed infinitely many times by the trajectory, with different incident angles. 
Moreover, as $s_J$ runs essentially only near the potential walls,
this description mimics again the reduction of the flow to a discrete 
map \cite{Berger,Aquila2}.
In such a way, both the maps and the Riemannian
pictures, base their results about stability on substantially equivalent approximations. 
In conclusion, an unapproximated method, which can give a full
description of the eventual instability of the Bianchi IX dynamics, is still lacking.\\
 Aim of this work is to provide such a description in the framework of the geometrodynamical 
approach to Chaos, generalizing the Jacobi description of motion in order to
avoid any singular behaviour (see \cite{Sigrav96}). 
We explicitly analyze the stability properties by numerically integrating 
the geodesics and geodesic deviation equations in this more general geometrical setting.\\
The generalization of Riemannian manifolds is accomplished using the theory of 
Finsler Spaces \cite{Rund},
in which the line element, {\it i.e.} the invariant 
time parameter, and all the geometrical quantities (in particular the metric and the 
curvatures) depend
on both coordinates and velocities.
The Finsler geometrodynamical approach to Chaos has been already exploited in other 
dynamical systems. 
In \cite{Aquila2}, we compared the tangent dynamics with the geodesic deviation equations 
of the Jacobi and Finsler metrics to highlight differencies and analogies, taking as 
paradigmatic
classical dynamical system, the H\'enon-Heiles Hamiltonian. This system has been widely 
investigated in \cite{HeHe}, where we find a strong evidence of correlation between
geometric properties of the Finsler manifold and regular or chaotic behaviour
of single orbits.
In the Finsler metric, the natural definition of the invariant interval of time is given by
\be
ds\equiv ds_F \doteq {\cal L}(q^i, \dot q^i) dt \quad (i=1,\dots ,N) \ ,      
\label{dsF}
\ee
where ${\cal L} (q^i, \dot q^i)$ is the Lagrangian of the system and $N$ is the number 
of degrees of freedom of the system. In \cite{PRE97} we developed
the mathematical formalism to analyze the instability properties of very general 
Lagrangian systems 
with $N$ degrees of freedom and performed some preliminary steps towards a 
gauge-invariant description of instability for the Bianchi IX dynamics. 
In order to consider the definition (\ref{dsF}) as a {\sl good} one and to
derive from it a metric tensor, the Lagrangian must fulfill, among others, the 
condition to be sign definite (e.g., positive). 
This can be obtained quite simply, remembering that it is always possible to add to the
Lagrangian the total derivative of a function of coordinates, 
without changing the dynamics.
As the condition ${\cal L}>0$ can be fulfilled along all the trajectory, the  
interval $ds_F$ never vanishes and 
all the singularities in the curvatures are avoided, also for peculiar dynamical 
systems as the Bianchi IX models, and we can {\sl geometrize} the full motion. 
The Finsler metric is given by 
\be
g_{\beta\gamma}(x^\alpha, x^{\prime\alpha}) = {1\over 2} 
{{\partial^2 \Lambda^2(x^\alpha, x^{\prime\alpha})}
\over{\partial x^{\prime \beta}\partial x^{\prime \gamma}}} \ ,
\label{metrica}
\ee
where $\Lambda$ is the homogeneous Lagrangian
\be
\Lambda (x^\alpha, x^{\prime\alpha}) = 
{\cal L} \left(t, x^i, {{x^{\prime i}}\over {x^{\prime 0}}}\right) 
x^{\prime 0}, \ (\alpha,\beta,\gamma = 0,\dots , N) ,
\ee 
$x^0 = t$ and $x^{\prime \alpha} = dx^\alpha/ds_F$.
As we see, the metric and all the other geometrical quantities depend on both 
coordinates and velocities.
The mathematical tool to investigate the stability, in the geometrical framework, is 
given by the geodesic deviation equations which, in local coordinates, read
\be 
{\displaystyle {{\nabla}\over{ds}}\left({{\nabla z^\alpha}\over {ds}}\right)}
= -R^\alpha{}_{\beta\gamma\delta} {dx^\beta\over ds} z^\gamma {dx^\delta\over ds}
= - {\cal H}^\alpha{}_\gamma z^\gamma \ , \label{EDG}  
\ee 
where $z^\alpha$ is the perturbation to a geodesic, $ \nabla/ds $
is the total (covariant) derivative along the geodesic,
$R^\alpha{}_{\beta\gamma\delta}$ is one of the curvature tensors defined in Finsler 
Geometry\cite{Rund} and ${\cal H}$ is the so-called {\sl stability tensor} 
\cite[B)]{commento1}.\\
The dynamical behaviour of the system is led by this tensor, whose eigenvalues,
$\{\lambda_\alpha\}$, are the sectional
curvatures and whose trace ${\sf Ric}$ is the analogous of the 
Ricci curvature (along the geodesic) in Riemannian geometry.
In analogy with the Lyapunov exponent, we define an instability exponent 
\be
\delta_I = \lim_{s \rightarrow \infty}\ \lim_{z(0) \rightarrow 0}
\left[{\displaystyle 
{ {1\over s}\ \ln {{z(s)\over z(0)}} } }\right]  \ , \label{exp_inst}
\ee
which now turns to be a quantitative
indicator, invariant under change of coordinates $x^\alpha =(t, x^i)$, 
giving a measure of a possible exponential growth of the perturbation $z(s)$, 
defined as the {\sl natural norm} on the manifold,
$ z^2 = g_{\alpha\nu}\, z^\alpha z^\nu\ ,$
in terms of the geodesic time parameter.
In order to grasp some qualitative informations on the stability of the flow, 
some simplified
approximated versions of the $(N+1)-$equations (\ref{EDG}) can be derived, 
as the ones obtained
substituting the covariant derivatives with ordinary ones and the
rhs of eq.(\ref{EDG}) with the $N$ non trivial eigenvalues of ${\cal H}$:
\be
{d^2 z_{(i)} \over ds^2} + \lambda_i z_{(i)} = 0 \ \ (i=1,\dots , N) \ . \label{sect}
\ee
When $N\gg 1$, a further simplification is attained summing the $N$ equations (\ref{sect}),
obtaining the single scalar equation:
\be
{d^2 z \over ds^2} + {{\sf Ric}\over N} z = 0 \label{Ricci}
\ee
which has been successfully applied, for some $N$-body systems \cite{commento1,Aquila1}.\\
In the case of the Bianchi IX dynamics we have considered the Lagrangian
\be
{\cal L} = K - U(\alpha,\beta_+,\beta_-) \label{Lag}
\ee
where $2K = \dot\beta_{+}^2 + \dot\beta_{-}^2 - \dot\alpha^2$, 
$U$ is the minisuperspace potential \cite{Misner}, 
$3\alpha={\rm ln} (abc)$, $6\beta_-= \sqrt{3}\cdot {\rm ln}(a/b)$ and 
$6\beta_+= {\rm ln} (ab/c^2)$, $a$, $b$ and $c$ are the scale factors
and the dot means differentiation with respect to
the time parameter $d\tau=dt/abc$, defined in terms of the {\sl cosmological proper time} $t$.
The time component of the field equations imposes the constraint ${H} = K + U = 0$, whose 
fulfillment must be checked carefully.
We numerically integrated the equations of motions, using two different methods: a 
fourth order Runge-Kutta scheme and a symplectic algorithm, both with a variable time step.
The quantity 
$ \Delta = |{H}|/(\dot\beta_{+}^2 + \dot\beta_{-}^2 + \dot\alpha^2)$ 
has been monitored in order to test the accurancy
of the two algorithms \cite{notaDelta} and we find that it never exceeds the value $10^{-5}$.
In fig.\ref{fig1} the three non vanishing sectional curvatures and the Ricci curvature 
have been reported versus $s_F$. All them evolve towards zero but some peaks are
present in coincidence with the bounces on the potential walls. 
Note that $\lambda_{1,2}$ are not sign definite but both are negligible compared to 
$\lambda_3$, which is instead always positive. 
Therefore, in constrast with the Jacobi metric, also the Ricci curvature is always positive. 
It has been already remarked that in some cases equations (\ref{sect}) 
and (\ref{Ricci}) describe correctly the qualitative features of dynamics, \cite{PRE97},
and some authors \cite{Biesiada,Sota} explicitly considered them. 
So, in fig.\ref{fig2} we plot the instability exponents 
calculated from these equations. Note that {\sl all} the $\delta_I$ go to zero,
confirming the absence of any global exponential instability. 
However, to be {\sl rigorous}, we also considered the exact geodesic deviation equations 
in the following way. As the two variables $\tau$ and $\alpha$ are monotonic 
functions\cite{Rugh} of $s_F$, the possibly unstable behaviour 
cannot be ascribed to them, so we restrict ourselves to consider the growth of the projection
$\tilde z$ of the perturbation $z$ in the plane $\beta_+\beta_-$, 
which is also the physically meaninguful {\it space} in this context,
\be
{\tilde z}^2 = g_{ij} z^i z^j \quad (i,j = \beta_+,\beta_-) \ . \label{tildez}
\ee
These are the only variables
which can give rise to an exponential instability, as they describe the anisotropy
of the model. The metric in this submanifold is moreover positive definite.
In fig.\ref{fig3} we plot the instability exponent for $\tilde z$ for several 
initial conditions.
As in the case of fig.\ref{fig2}, for all the conditions we consider, there is a clear 
absence of any exponential divergence.
However, it must be stressed again that, in correspondence of the bounces, 
the $\delta_I$ tends always to increase (see the inset in fig.\ref{fig3}), 
though the amount of this rise reduces during the evolution; 
correspondingly, the $s_F-$interval between two successive bounces increases, roughly
exponentially, so that the instability exponents undoubtedly decrease to zero as 
$\delta_I\propto s_F^{-1}$.
From the {\sl strict definition} of Chaos, we should then conclude that
the system is not chaotic due to the absence of a global exponential divergence of nearby
trajectories. However, the local sharp increase of the $\delta_I$'s
reveals that a different kind of impredictable behaviour, 
confined in very short time intervals, could be present, as also outlined by Contopoulos
et al. \cite{Contopoulos}, though within a very different approach.
The unpredictability in the duration of each long era and in the number of Kasner epochs for
each era, is another signature of the stochastic nature of the Bianchi IX dynamics, 
which reflects itself in the sharp rises, due to the {\sl  chaotic
reflections} from the potential walls. 
It is obvious that stochasticity depends on the definition adopted and also on the 
observable chosen.\\
In summary, we presented a gauge invariant approach to Chaos for Bianchi IX dynamics,
without assuming any kind of approximations.
Our results reveal the absence of any global chaotic behaviour even if they suggest
the occurrence of a chaotic scattering phenomenon.
Further details and a wider discussion on the subject will be presented elsewhere.


\begin{figure}
\caption{Evolution of the sectional curvatures $\lambda_1$, 
$\lambda_2$ and $\lambda_3$ and of the Ricci curvature per degree of freedom, 
${\sf Ric}/3$, for the initial conditions $\beta_+=0.33$,  $\dot\beta_+=-0.23$,  
$\beta_-=0$,  $\dot\beta_-=-0.64$ and $\alpha=-0.3$. 
Note that only $s_F$ is plotted in log scale.}
\label{fig1}
\end{figure}
\begin{figure}
\caption{Evolution of the instability exponents $\delta_{\lambda_1}$, 
$\delta_{\lambda_2}$, $\delta_{\lambda_3}$  and 
$\delta_{{\sf Ric}/3}$  for the approximated geodesic deviation equations 
(\ref{sect}) and (\ref{Ricci}). The initial conditions are the same as in fig.\ref{fig1}.}
\label{fig2}
\end{figure}
\begin{figure}
\caption{Evolution of the instability exponents  $\delta_{\beta_+\beta_-}$ 
defined for $\tilde z$ in equation (\ref{tildez}), for several initial conditions. 
In the inset it is plotted the curve 
corresponding to the same initial condition of figs.\ref{fig1} and \ref{fig2}.}
\label{fig3}
\end{figure}

\begin{references}
\bibitem[*]{byline} Also at {\sf I.N.F.M.} Unit\`a di Ricerca di Parma;\\
e-mail: mariateresa.dibari@fis.unipr.it

\bibitem[\dagger]{byline} Also at {\sf I.N.F.M.} Unit\`a di Ricerca di Roma 2;\\
e-mail: cipriani@axtov1.roma2.infn.it

\bibitem{Hobill_book}
For a recent review on the subject see
D. Hobill, A. Burd and A. Coley {\sl Deterministic Chaos in General Relativity}, 
Plenum Press (1994) and references therein. 
An older but very interesting review on Chaos in General Relativity is given 
by J. D. Barrow, Phys.  Rep., {\bf 85}, 1 (1982).

\bibitem{Prigo}
The {\it internal time} is the time measured by a {\it natural clock}, associated with the
dynamics. For a discussion on the definition of ``time" near the singularity 
in a cosmological context, see 
C. W. Misner, K. Thorne, J. A. Wheeler, {\it Gravitation}, Freeman, San Francisco (1977);
C. M. Lockhart, B. Misra and I. Prigogine, Phys. Rev. D {\bf 25}, 921 (1982).

\bibitem{Berger}
B. K. Berger, Gen. Rel. Grav. {\bf 23}, 1385 (1991);
A. B. Burd, N. Buric, R. K. Tavakol, Class. Quantum Grav. {\bf 8}, 123 (1991).

\bibitem{Bel}
V. A. Belinskii, I. M. Khalatnikov and E. M. Lifshitz, Adv. \ Phys. {\bf 19}, 525 (1970); 
{\bf 31}, 639 (1982).

\bibitem{Barrow}
J. D. Barrow, Phys. Rev. Lett. {\bf 46}, 963 (1981);
D. F. Chernoff and J. D. Barrow, Phys. Rev. Lett. {\bf 50}, 134 (1983);
I. M. Khalatnikov, E. M. Lifshitz, K. M. Khanin, L. N. Shchur, Ya G. Sinai, J. Stat. Phys. {\bf 38}, 97 (1985).

\bibitem{Cornish}
N. J. Cornish and J. Levin, Phys. Rev. D {\bf 55}, 7489 (1997). 

\bibitem{Misner} 
C. W. Misner, Phys. Lett. {\bf 22}, 1071 (1969). 

\bibitem{maps}
A. J. Lichtenberg and M. A. Lieberman, {\sl Regular and Stochastic Motion},
Springer-Verlag (1983).

\bibitem{Chitre}
D. M. Chitre {\it Ph.D. Thesis}, Univ. of Maryland (1972).

\bibitem{Anosov} 
D. V. Anosov, Proc.Steklov Inst. Math. {\bf 90}, 1 (1967).

\bibitem{Szydlowsky}
M. Szydlowski and A. Krawiec, Phys. Rev. D {\bf 47}, 5323 (1993).

\bibitem{commento1}
Several systems with many degrees of freedom have been widely studied with 
Jacobi and Eisenhart metrics: {\bf A)} Pettini, Phys. Rev. E {\bf 47}, 828 (1993); 
L. Casetti, R. Livi and M. Pettini, Phys. Rev. Lett. {\bf 74}, 375 (1995);
{\bf B)} P. Cipriani, {\it Ph.D.Thesis}, Univ. of Rome ``La Sapienza", in italian (1993); 
P. Cipriani and G. Pucacco, Il Nuovo Cimento B {\bf 109} n.3, 325 (1994); 
see also \cite{Aquila1}.

\bibitem{Biesiada}
M. Biesiada and S. E. Rugh, preprint gr-qc/9408030 (1994).

\bibitem{Burd}
A. Burd and R. Tavakol, Phys. Rev. D {\bf 47}, 5336 (1993).

\bibitem{Aquila2}
M. Di Bari and P. Cipriani, Planetary and Space Science, in press (July 1998).

\bibitem{Sigrav96}
M. Di Bari and P. Cipriani
in {\sl Proc. of 12$^{th}$ Italian Conference on General Relativity 
and Gravitational Physics}, World Scientific, 407 (1997).

\bibitem{Rund}
H. Rund (1959), {\it The differential Geometry of Finsler Spaces}, Springer-Verlag (1959);
M. Matsumoto, {\sl Foundations of Finsler Spaces}, Kaiseisha Press - Saikawa, 
Otsu, Japan (1986).

\bibitem{HeHe}
P. Cipriani and M. Di Bari, submitted to Phys. Rev. Lett. (1998).

\bibitem{PRE97}
M. Di Bari M., {\it Ph.D.Thesis}, Univ. of Rome ``La Sapienza", in italian (1996);
M. Di Bari {\it et al.}, Phys. Rev. E {\bf 55}, 6448 (1997).

\bibitem{Aquila1}
For some considerations on the two kind of approximations, 
see P. Cipriani and M. Di Bari, Planetary and Space Science, in press (July 1998).

\bibitem{notaDelta}
Taking into account the asymptotic behaviour of $|K|$ and $|U|$, it is clear that a (small)
upper limit on $\Delta$ rather than on $|H|$ itself, gives a much more sensitive
check on the constraint.

\bibitem{Sota}
Y. Sota, S. Suzuki, K. Maeda, Class. Quantum Grav. {\bf 13}, 1241 (1996).

\bibitem{Rugh}
S. E. Rugh, B. J. T. Jones, Phys. Lett. A {\bf 147}, 353 (1990).

\bibitem{Contopoulos}
G. Contopoulos, B. Grammaticos and  A. Ramani, J. Phys. A {\bf 28}, 5313 (1995).
\end{references}
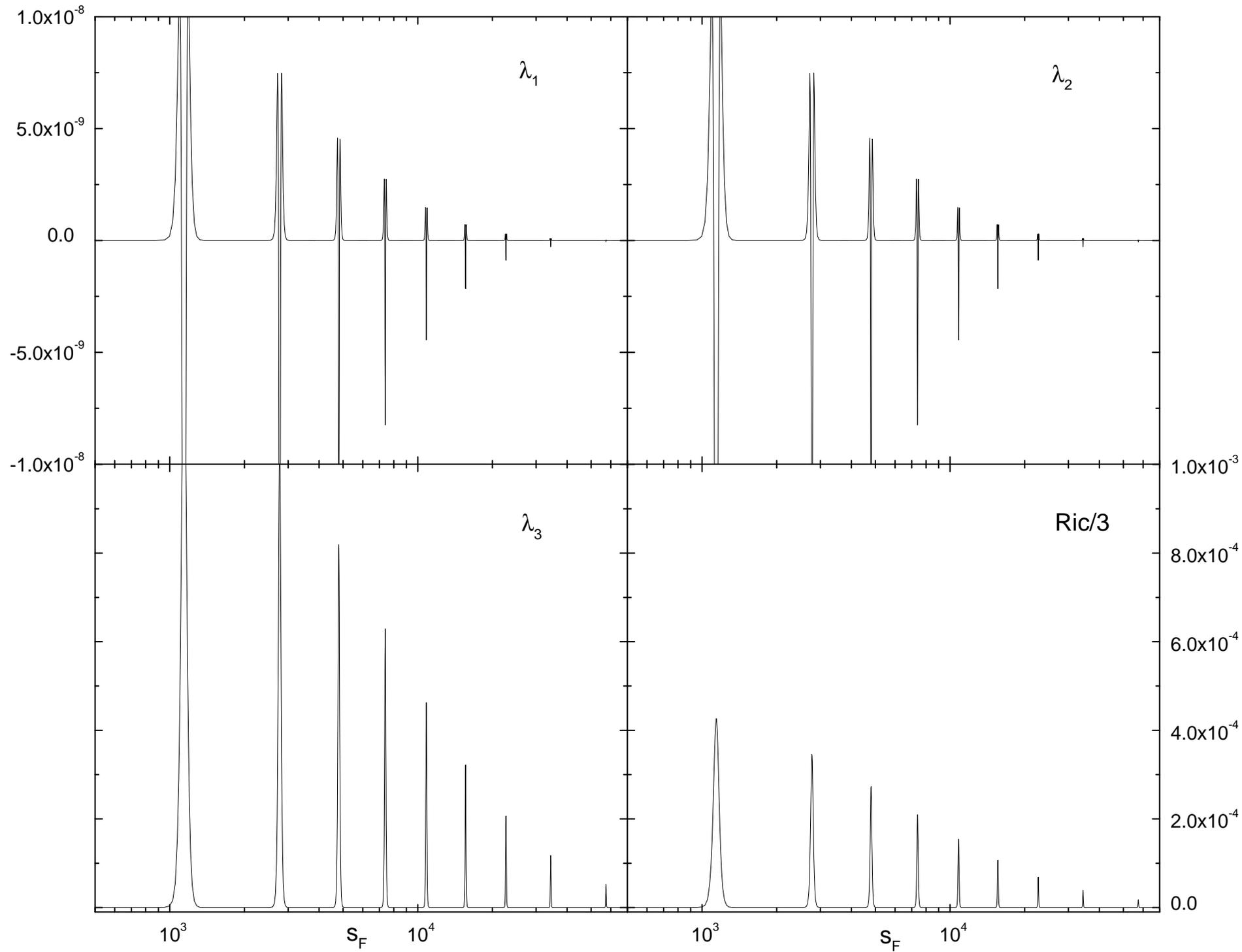
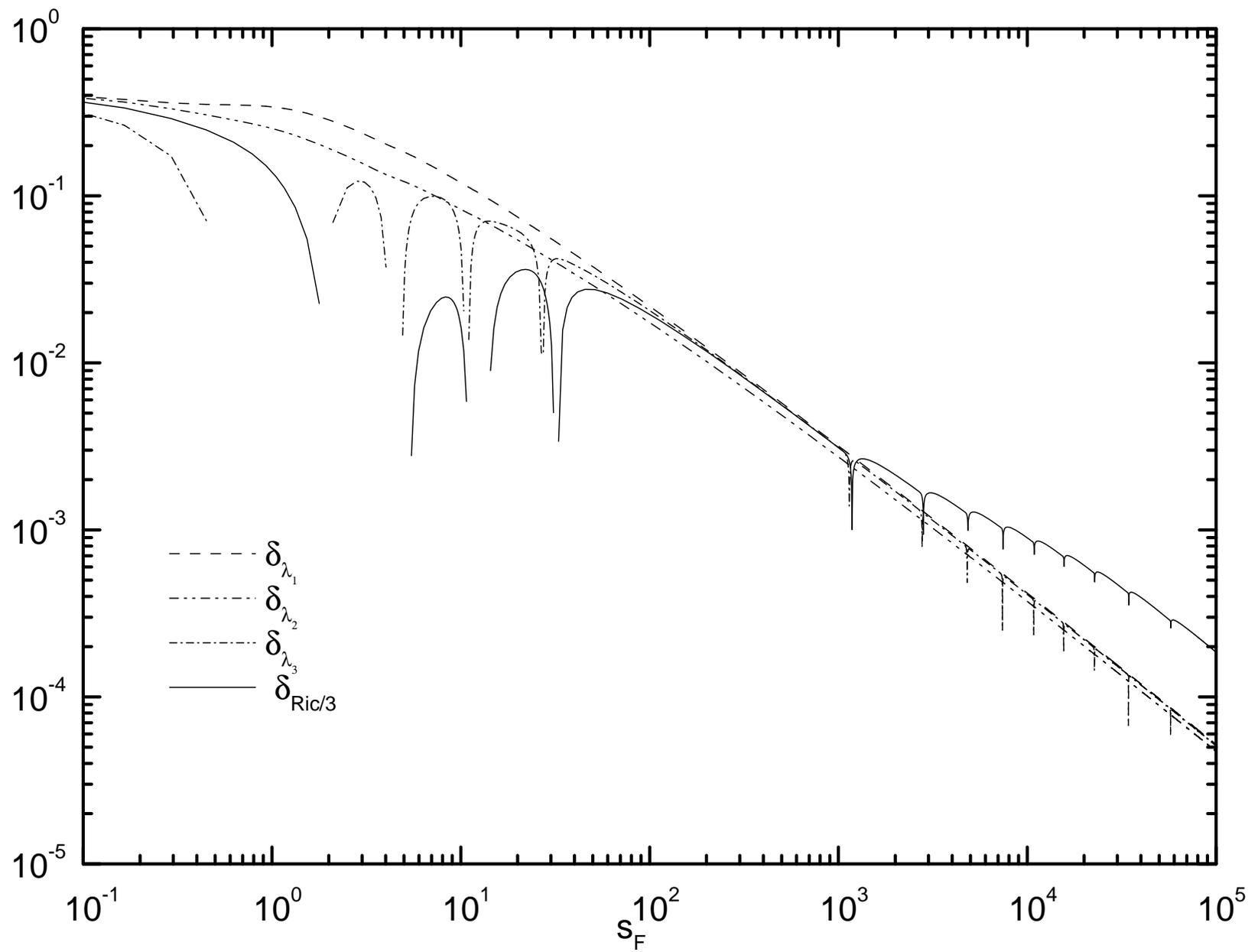
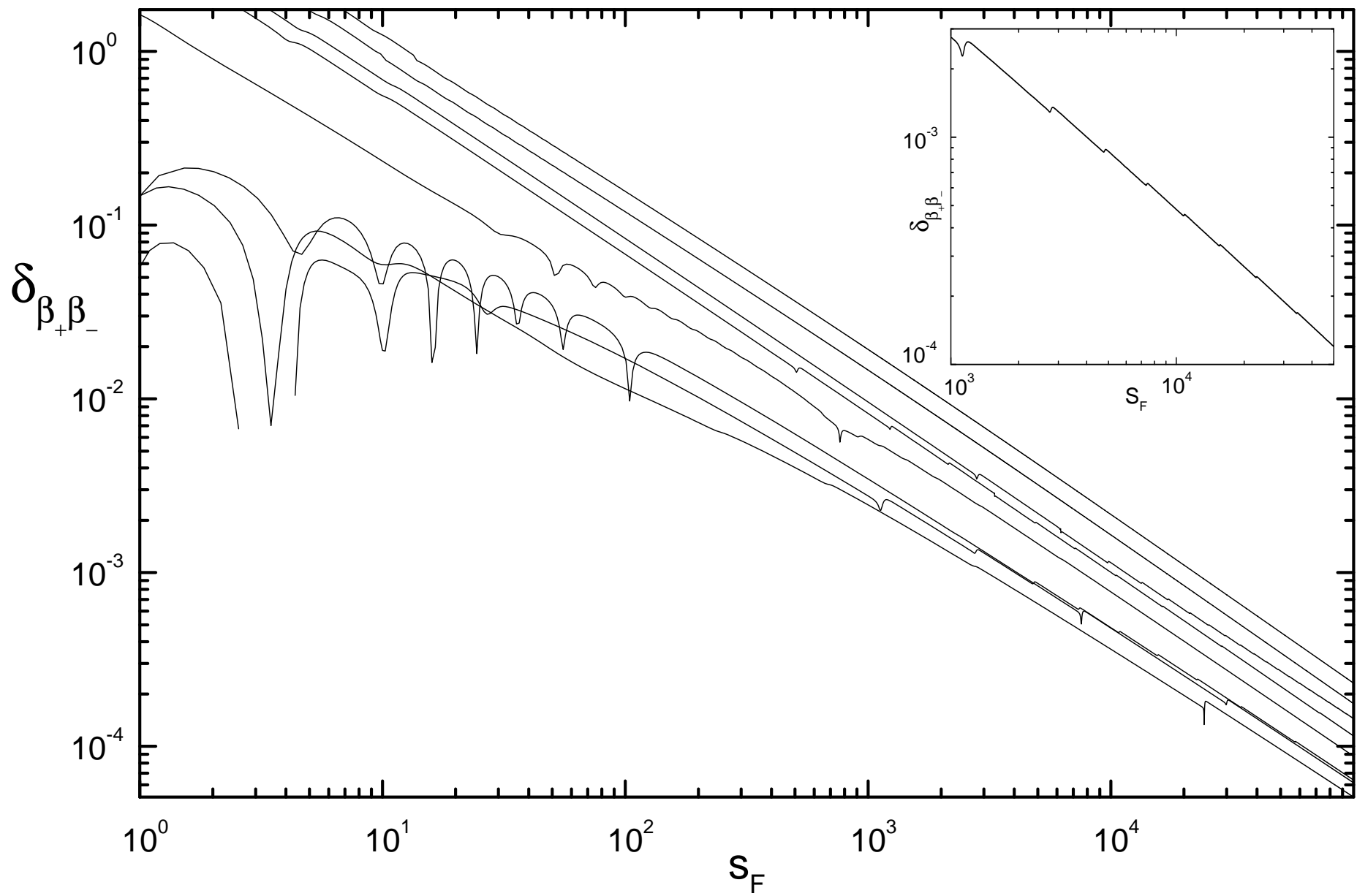
\end{document}